\documentstyle{article}
   \begin{document}
   \setlength{\baselineskip}{15pt}
   \title{Non polynomial vector fields under the Lotka--Volterra normal form}
   \author{ Benito Hern\'{a}ndez--Bermejo  \and
            V\'{\i}ctor Fair\'{e}n  }
   \date{}

   \maketitle
   \mbox{}

   {\em Departamento de F\'{\i}sica Fundamental,
   Universidad Nacional de Educaci\'{o}n a Distancia. Aptdo. 60141,
   28080 Madrid (Spain). E--mail vfairen@uned.es .}

\mbox{}

   \begin{abstract}
We carry out the generalization of the Lotka-Volterra embedding to flows not
explicitly recognizable under the Generalized Lotka-Volterra format.
The procedure introduces appropiate auxiliary variables, and it is shown
how, to a great extent, the final Lotka-Volterra system is independent of
their specific definition. Conservation of the topological equivalence
during the process is also demonstrated.
   \end{abstract}
\pagebreak

\section{Introduction}
The concept of normal form is frequently used in mathematical physics
to describe peculiar forms of mathematical objects for which
the properties of a large class of them are particularly evident
and easy to analyze. They are thus especially useful in the context
of classification of those mathematical objects, and found
in matrix theory (Jordan form) \cite{lips}, catastrophe theory \cite{gil}, 
bifurcation theory \cite{craw}, etc.

Contrarily to the theory of linear ODE's there is no normative approach
in its nonlinear counterpart through which we could relate the properties
of the structure of the equations themselves to the properties of their
corresponding solutions. The problem lies in the lack of unifying structures
for nonlinear representations.
There has been, however, a great deal of fruitful work devoted to this subject,
directed towards the elaboration of recasting procedures of general nonlinear systems
into standard simple forms, which are amenable to systematic analysis.
In the early thirties Carleman  \cite{car} showed how a nonlinear system of 
ODE's could be transformed into an infinite-dimensional linear system. Kowalski
and Steeb  \cite{kys} have reviewed the theory and applications of this embedding,
while Cair\'{o} and Feix  \cite{cyf} have recently shown the interest of the technique
in the study of invariants of the motion. An alternative approach has been
suggested by Kerner \cite{ker}, in terms of an embedding into a Riccati format.
This scheme can be plugged into the classification and analysis of
quadratic ODE's by means of non-associative algebras \cite{xxx}. The Riccati
embedding has also been successfully used in time-saving, high precision
numerical codes \cite{vfea,pic} and in nonlinear model building and prediction in chaos \cite{lop}.
Finally, we can mention the S-systems formalism, originally devised in the
context of representation of nonlinear biochemical schemes, but which
has proved to be a good framework for the recasting and solution of
nonlinear differential systems \cite{voit}.

The central idea behind these unifying representations is that
they should constitute the framework for
building a method of classification based on powerful algebraic techniques,
differing from the more traditional
geometric methods customarily
used in the qualitative theory of differential equations, especially
in high dimensional systems, in which we can lose the intuitive benefits
of the geometric approach. The gain in structural simplicity is bought
at the price of an increase in the dimension of the system, but this cost
will become negligible as soon as further progress is made on the subject.

In this context, Brenig and Goriely
\cite{bre1,bre2} have recently introduced the concept of Generalized Lotka-Volterra
(GLV) form:
\begin{equation}
   \dot{x}_i = x_{i}(\lambda _{i} + \sum_{j=1}^{m}A_{ij}\prod_{k=
      1}^{n}x_{k}^{B_{jk}}) , \;\: i = 1 \ldots n , \; m \geq n.
   \label{eq:glv}
\end{equation}
where $A$ and $B$
are $n \times m$ and $m \times n$ matrices,
respectively. The interesting point here lies in the fact that the properties
of system (1) are assotiated to those of algebraic objects as are
matrices $\lambda$, $A$ and $B$: the family of systems (1) is split
in classes of equivalence according to the values of
the products $B\cdot A$ and $B\cdot \lambda$. Accordingly,
certain prescribed forms of those products define representative
systems of the class for which an integrability analysis can be
systematically carried out \cite{gori}.
The classical Lotka-Volterra (LV) system is one of those forms.
If matrix $B$ is of rank $n$, (\ref{eq:glv}) can be
embedded into the following m-dimensional system:
\begin{equation}
   \dot{z} _{\alpha} = \lambda '_{\alpha}z_{\alpha} + z_{\alpha}\sum_{\beta = 1}^{m}
      A'_{\alpha \beta} z_{\beta} , \;\: \alpha = 1,\ldots ,m.
   \label{eq:lvf}
\end{equation}
where each one of the
Lotka-Volterra variables $z_j$ stands for any one of the
quasimonomials:
\begin{equation}
   \prod_{k=1}^{n}x_{k}^{B_{jk}} \; , \;\: j = 1, \ldots , m \; ,
   \label{eq:qm}
\end{equation}
in (\ref{eq:glv}), while $A ' = B \cdot A$ and
$\lambda ' = B \cdot \lambda$.

The simplicity and ubiquity of the Lotka-Volterra equations (\ref{eq:lvf})
has made them especially attractive and amenable to systematic analysis
by algebraic techniques \cite{alm}. They are algebraically related to
the replicator equations \cite{hys}, of primary importance in mathematical
biology and for which we have fairly general results, and also
to the rapidly growing theory of neural nets \cite{noon}.
Moreover, the Lotka-Volterra equations can be straightforwardly
embedded within a connectionist representation of dynamical systems,
in terms of digraphs and nodes, an approach which opens exciting
prospects for a near future \cite{farm}.

The purpose of the present letter is to show
how a fairly general class of differential
systems, apparently not covered by the GLV form, can be easily rewritten in
terms of (\ref{eq:glv}) by means of the introduction of
suitable auxiliary variables. We may recall all vector fields
containing elementary functions: rational, exponential, etc.
Those systems can then be recast in terms of an equivalent Lotka-Volterra
system, thus providing, as mentioned before, a possible route from the 
traditional view of the
field --as a potpourri of structureless and apparently unrelated systems--
to a new emerging unifying one.

\section{Embedding into the GLV form}
We shall consider a system of the general form:
   \[ \dot{x}_{s} = \sum_{i_{s1}, \ldots ,i_{sn},j_{s}} a_{i_{s1} \ldots
        i_{sn} j_{s}} x_{1}^{i_{s1}} \ldots x_{n}^{i_{sn}}f(\bar{x})^{j_{s}} \]
   \begin{equation}
         x_{s}(t_{0}) = x_{s}^{0}, \; \: s=1, \ldots , n,
     \label{eq:ini}
   \end{equation}
where $f(\bar{x})$ is some scalar function not reducible to the quasimonomial
form (\ref{eq:qm}) in terms of the $\bar{x}$
variables. All constants in (\ref{eq:ini}) are assumed
to be real. There is no loss of generality in what is to follow in
elaborating on systems with a single non quasimonomial function, as in
(\ref{eq:ini}).
The addition to the scheme of other appropiate functions, if eventually
needed, does not alter the method we are going to develop.

Now, we additionally assume that $f(\bar{x})$ is such that its partial derivatives can
be expressed in the following form:
\begin{equation}
   \frac{\partial f}{\partial x_{s}} = \sum_{e_{s1},\ldots ,e_{sn},e_{s}}
      b_{e_{s1} \ldots e_{sn} e_{s}}x_{1}^{e_{s1}} \ldots x_{n}^{e_{sn}}f(\bar{x})^{e_{s}}
   \label{deriv}
\end{equation}
All constants are again real numbers.
The possibility of dealing with an $f(\bar{x})$ whose derivatives do not
verify (\ref{deriv})
does not actually affect the generality of the following considerations.
This problem has already been
analyzed~\cite{ker}; it can be shown that any function satisfying
a finite order differential equation can be reduced to a first order
polynomial system by means of a method that consists in assigning
new variables to the nonpolynomial terms and differentiate them to find
their differential equations. The process is repeated successively
until we finally reach an expanded
polynomial differential first order system equivalent to the initial
equation (suitable initial conditions must be taken into account). For
example, if $f(x) = \sin x$, the first derivative is not polynomial in $x$ and $f$
\begin{equation}
      \frac{df}{dx} = \sqrt{1-f^{2}}\: ,
   \label{seno}
\end{equation}
but the sine function is also the solution of
\begin{equation}
      \frac{d^{2}f}{dx^{2}} + f = 0 .
      \label{seno2}
\end{equation}
The above method would introduce a new variable $q = \mbox{d}f/\mbox{d}x$
reducing (\ref{seno2}) to:
\[ \frac{df}{dx} = q , \]
\[ \frac{dq}{dx} = -f . \]
There is then no loss of generality in assuming (\ref{deriv}).

The choice of $f(\bar{x})$ in (\ref{eq:ini}) is certainly ambiguous
given the fact that any other function of the form
     \[ f_{k, \vec{l}}(x) = f^{k} \prod_{s=1}^{n} x_{s}^{l_{s}} \;\: , \]
with $k \neq 0$, will also preserve the format (\ref{eq:ini}).
Moreover, expression (\ref{deriv}) may not be unique for
a given $f(\bar{x})$. For example, to
\[ f(x) = \frac{x^{2}}{1+x^{2}}, \: \; (n=1) \]
we may associate a countable family of possible quasipolynomial
representations of the derivative:
\[ \frac{df}{dx} = 2x^{-2i-3}f^{i+2}(1+x^{2})^{i}, \: \; i \in \cal{N} \]
In the development to come, we shall henceforth proceed for one given
selection
in (\ref{eq:ini}) and (\ref{deriv}) of both $f(\bar{x})$ and the form of
its derivatives.

The procedure to transform (\ref{eq:ini}) and (\ref{deriv}) into a GLV system is
then straightforward. It is carried out by introducing an additional
variable in the form
\begin{equation}
   y = f^{q}\prod_{s=1}^{n} x_{s}^{p_{s}}, \:\; q \neq 0 \; ,
   \label{eq:cam}
\end{equation}
with real exponents $q, \: p_{s}$. The set (\ref{eq:cam}) of all possible
new variables includes $y = f(\bar{x})$ as a special (and simplest) element.
The transformations which map $f(\bar{x})$ onto any other element of the set,
\[ \xi (\vec{p},q): f \rightarrow f^{q} \prod_{s=1}^{n} x_{s}^{p_{s}} , \]
constitute a $(n+1)$ parameter non conmutative Lie group, with the
composition
\[ \xi (\vec{p}_{1},q_{1}) \circ \xi (\vec{p}_{2},q_{2}) = \xi
(\vec{p}_{1} + q_{1} \vec{p}_{2}, q_{1} q_{2}) \]
as inner operation. Thus, the set of functions (\ref{eq:cam}) generated
from $f$
make up an equivalence class, where for every two elements $f_{i}$
and $f_{j}$ there is a group element $\xi _{ij}$ which transforms one into the
other:
\[ \xi _{ij} : f_{i} \rightarrow f_{j} \]

The introduction of the auxiliary variable (\ref{eq:cam}) leads to the
following system for the original variables:
\begin{equation}
   \dot{x}_{s} = \left[ x_{s}\sum_{i_{s1}, \ldots ,i_{sn},j_{s}} a_{i_{s1} \dots
      i_{sn} j_{s}}y^{j_{s}/q}\prod_{k=1}^{n} x_{k}^{i_{sk}- \delta _{sk} -
      j_{s}p_{k}/q} \right]
   \label{eq:x}
\end{equation}
for $s = 1, \ldots ,n$. As usual, $ \delta _{sk} = 1$ if $s=k$, and 0 otherwise.
For the new variable (\ref{eq:cam}) we obtain
\[
   \dot{y} = \sum_{s=1}^{n} \frac{\partial y}{\partial x_{s}}
         \dot{x} _{s} =
   y \left[ \sum_{s=1}^{n} \{ p_{s}x_{s}^{-1}\dot{x}_{s} + \right. \]

\begin{equation}
   \left. + \sum_{i_{s \alpha },j_{s},e_{s \alpha },e_{s}}a_{i_{s \alpha },j_{s}}
   b_{e_{s \alpha }
   e_{s}} q y^{(e_{s}+j_{s}-1)/q}\prod_{k=1}^{n}x_{k}^{i_{sk}+e_{sk}
   + (1-e_{s}-j_{s})p_{k}/q} \} \right]
   \label{eq:y}
\end{equation}
where $\alpha = 1, \ldots ,n $. An appropiate initial condition $y(t_{0})$
must also be included (this will be assumed whenever a new
variable is introduced). With (\ref{eq:x}) and (\ref{eq:y}) the reduction of
system (\ref{eq:ini}) to the GLV format is achieved.

\section{Embedding into the LV form}
In order to prove our assertions on (\ref{eq:x})--(\ref{eq:y}) we shall find of
interest to use some known results from  Brenig and Goriely~\cite{bre1,bre2},
which we briefly recall. A general GLV system (\ref{eq:glv})
is formally invariant under quasimonomial transformations
\begin{equation}
   x_{i} = \prod_{k=1}^{n} \hat{x} _{k}^{C_{ik}} , \;\: i=1,\ldots ,n
   \label{bec}
\end{equation}
for any invertible matrix $C$. The matrices $B, A$ and $\lambda$ change to
$\hat{B}=B \cdot C, \hat{A}=C^{-1} \cdot A$ and $\hat{\lambda}=C^{-1}
\cdot \lambda$, respectively, but the GLV format is preserved.
These transformations define the important concept of Brenig's equivalence
classes (BEC), which consist of all the GLV systems related through
transformations (\ref{bec}). In such BEC's, the products $B \cdot A$ and $B
\cdot \lambda$
are invariants of the class. As mentioned in the introduction these invariants
define the LV matrices, $A'$ and $\lambda '$, in (\ref{eq:lvf}). Then, the LV
matrices are unique for a given BEC. This summarizes earlier results from
Brenig and Goriely.

We are now in a position to introduce our two preliminary propositions
concerning general properties of the BEC's.

\newtheorem{th}{Theorem}
\newtheorem{co}{Corollary}
\newtheorem{lm}{Lemma}
\begin{lm}
 The quasimonomials of a GLV system are also invariants of the BEC to which
 the GLV system belongs.
\end{lm}

{\bf Proof:}

If we call $D = C^{-1}$ the i-th new quasimonomial $1 \leq i \leq m$ will be
\[ \prod_{j=1}^{N}\hat{x}_{j}^{\hat{B}_{ij}} =
   \prod_{j=1}^{N} \left[ \prod_{k=1}^{N}x_{k}^{D_{jk}} \right] ^{\hat{B}_{ij}} =
   \prod_{k=1}^{N}x_{k}^{ \left( \sum_{j=1}^{N}\hat{B}_{ij}D_{jk} \right) } =
   \prod_{k=1}^{N}x_{k}^{B_{ik}}  \]
The quasimonomials are, then, conserved in the class of equivalence ({\em q.e.d.\/}) .

As was mentioned in section 1, the quasimonomials of the GLV system are
precisely the LV variables. Thus, since both the LV matrices and variables
are the same for the whole BEC, we have proved the following:
\begin{co}
 All GLV systems belonging to the same BEC embed into a single initial--value
 problem, defined by a unique LV system and initial condition on the
 quasimonomials (\ref{eq:qm}).
\end{co}

We are then ready to focus our attention on the generic GLV system
(\ref{eq:x})--(\ref{eq:y}). It is clear that different systems are obtained
for distinct
choices of the auxiliary variable (\ref{eq:cam}). We shall first demonstrate
that all these systems are part of one BEC, that is:
\begin{th}
 For any auxiliary variable (\ref{eq:cam}), all resulting GLV systems belong
 to the same BEC.
\end{th}

{\bf Proof:}

Given two different choices of auxiliary variables
\[ y_{i} = f^{q_{i}}\prod_{s=1}^{n} x_{s}^{p_{si}}, \:\; q_{i} \neq 0 \; ,
   \;\; i = 1,2 \;\: ,\]
the resulting sets of GLV variables will be, respectively,
$(x_{1}, \ldots ,x_{n}, y_{1})$ and $(x_{1}, \ldots ,x_{n}, y_{2})$. A
straightforward calculation shows that both sets of variables are connected
through a transformation of the kind (\ref{bec}), with $C$ given by:
\[ C = \left( \begin{array}{ccccc}
         1   &   0    & \ldots &   0    &   0    \\
         0   &   1    & \ldots &   0    &   0    \\
      \vdots & \vdots & \ddots & \vdots & \vdots \\
         0   &   0    & \ldots &   1    &   0    \\
      \alpha_{1} & \alpha_{2} & \ldots & \alpha_{n} & \beta
        \end{array} \right) \;\: , \]
where
\[ \alpha_{s} = p_{s2} - \beta p_{s1} \;\: , \beta = \frac{q_{2}}{q_{1}} \; . \]
Consequently, both systems are members of the same BEC ({\em q.e.d.}).

All systems complying the format (\ref{eq:x})--(\ref{eq:y}) are in
the same BEC: according to Lemma 1 they must thus possess identical
quasimonomials. This can be easily
checked if we rewrite such quasimonomials in terms of the original variables
$\bar{x}$ and $f(\bar{x})$. The corresponding equations for the $x_{s}$ are
\begin{equation}
   \dot{x}_{s} = x_{s} \left[ \sum_{i_{s1}, \ldots ,i_{sn},j_{s}} a_{i_{s1} \dots
      i_{sn}j_{s}}f^{j_{s}}\prod_{k=1}^{n} x_{k}^{i_{sk}-\delta _{sk}} \right]
   \label{eq:xf}
\end{equation}
with $s=1, \ldots ,n$. For the $y$ we obtain:
\begin{equation}
   \dot{y} = y \left[ \sum_{s=1}^{n} \{ p_{s}x_{s}^{-1}\dot{x}_{s} +
   \sum_{i_{s \alpha }j_{s}e_{s \alpha }e_{s}}a_{i_{s \alpha }j_{s}}b_{e_{s
   \alpha }e_{s}}qf^{e_{s}+j_{s}-1}\prod_{k=1}^{n}x_{k}^{i_{sk}+e_{sk}} \} \right]
   \label{eq:yf}
\end{equation}
where $\alpha = 1, \ldots ,n$. The quasimonomials, as
functions of $\bar{x}$, do not depend in any way on the definition of the
auxiliary variable (\ref{eq:cam}),
but only on constants from (\ref{eq:ini})--(\ref{deriv}).

From the previous line of argument the following conclusion holds:
\begin{th}
 The LV system (\ref{eq:lvf}) generated from (\ref{eq:ini}) is
 completely determined from the
 choices for $f(\bar{x})$ and the representation of its derivatives.
\end{th}

In order to finalize the analysis it is needful to verify the equivalence
between the solutions of the initial and final systems.
The conservation of the topology through the whole process which carries
(\ref{eq:ini}) into (\ref{eq:lvf}) is a necessary condition for
ensuring the equivalence between the initial and final
systems. A sufficient condition for this property (ref. 7, p. 22)
is the existence of a diffeomorphism conecting the initial and final phase
spaces. Since the dimension of the LV system is greater than that
of (\ref{eq:ini}) due to the succesive embeddings, such a diffeomorphism
should connect the phase space of (\ref{eq:ini}) and the manifold of ${\cal R}^{m}$
into which it is mapped.

The transformation embedding (\ref{eq:x})--(\ref{eq:y}) into (\ref{eq:lvf})
can be written as~\cite{bre2}:
\begin{equation}
   z_{\alpha} = \prod_{\beta =1}^{m} x_{\beta }^{{\cal B}_{\alpha \beta}} ,
      \; \: \alpha = 1, \ldots ,m,
   \label{eq:dif}
\end{equation}
where $x_{1}, \ldots ,x_{n}$ are the variables in (\ref{eq:ini}), $x_{n+1}
= y$ and $x_{\alpha} = 1$, for $\alpha = n+2, \ldots ,m$. $\cal{B}$ is the
matrix of the expanded GLV system (see~\cite{bre2} for details) which is
finally mapped onto the LV system.
Equation (\ref{eq:dif}) is mathematically a diffeomorphism: it is
obviously a differentiable and onto map. Thus, we only need to prove that
it is one to one. If we take logarithms in both sides of (\ref{eq:dif}):
\[ \left( \begin{array}{c} \ln z_{1} \\ \vdots \\ \ln z_{m} \end{array}
   \right) = \cal{B} \left( \begin{array}{c} \ln x_{1} \\ \vdots \\
   \ln x_{n+1} \\ 0 \\ \vdots \\ 0 \end{array} \right)  \]
$\mbox{Rank}({\cal B}) = m$ by construction. Then, for any two vectors
$\bar{x}_{1}$ and $\bar{x}_{2}$, ${\cal B} \ln (\bar{x}_{1}) \neq
{\cal B} \ln (\bar{x}_{2})$, unless
$\bar{x}_{1} = \bar{x}_{2}$. Thus the map is one to one and topology is
preserved by the succesive transformations.

A necessary requirement for the previous results to hold is that the vectors
$\bar{x}$ and $\bar{z}$ in (\ref{eq:dif}) must have strictly positive
entries. A necessary and sufficient condition for this
is that both the variables $x_{1}, \ldots ,x_{n}$ and the selected function
$f(\bar{x})$ in (\ref{eq:ini}) are positive. If this is the case, the LV
variables $z_{1}, \ldots ,z_{m}$ and the intermediate variables defined
along the process will be also strictly positive.

When this condition is not {\em a priori\/} satisfied, a phase-space
translation is to be performed:
\begin{eqnarray*}
   \bar{x}    & \longrightarrow & \bar{x}' - \bar{c} \\
   f(\bar{x}) & \longrightarrow & f'(\bar{x}' - \bar{c})-k
\end{eqnarray*}
In most cases arising in practice (for example with integer exponents),
the translation preserves the format (\ref{eq:ini}).

\section{Examples}
A first illustration is provided by an equation modelling the concentration
of an allosteric enzyme (ref. 8, p. 137)
\begin{equation}
  \frac{\mbox{d}x}{\mbox{d}t} = -x\frac{a + bx}{c + x + dx^{2}} \;\: ,
\end{equation}
where $a , b , c , d$ are positive real constants. If we exclude the
asymptotic state $x = 0$, then, for any transient, we can make the following
definition:
\[ f(x) = \frac{1}{c + x + dx^{2}} \; \: , \; \frac{\mbox{d}f}{\mbox{d}x}
        = -f^{2} - 2dxf^{2} \]
We now consider an auxiliary variable of the form (\ref{eq:cam})
\[  y = x^{p}f(x)^{ q} \; , \: \; q \neq 0  \; , \]
from which we shall obtain a GLV system. Independently of the concrete values
of $p$ and $q$, such GLV systems will always embed into an
unique LV system, as can be inferred from Theorem 2. To see this we introduce
$y$ as an explicit function of the
parameters $p$ and $q$. The resulting family of ($p$, $q$)-dependent
GLV systems is:
\begin{eqnarray*}
\dot{x} & = & x \left[ -ax^{-p/q}y^{1/q} -
   bx^{1-p/q}y^{1/q} \right]                     \\
\dot{y} & = & y \left[ -pax^{-p/q}y^{1/q}-
   pbx^{1-p/q}y^{1/q}+qax^{1-2p/q}y^{2/q}+
   \right.                                                      \\
        &   & \left. q(b+2da)x^{2-2p/q}y^{2/q}+2bdqx^{3-
   2p/q}y^{2/q} \right]
\end{eqnarray*}
There are then 5 different quasimonomials:
\[ x^{-p/q}y^{1/q} = f , \;\: x^{1-p/q}y^{1/q} =
   xf , \;\: x^{1-2p/q}y^{2/q} = xf^{2} , \]
\[ x^{2-2p/q}y^{2/q} = x^{2}f^{2} , \;\:
   x^{3-2p/q}y^{2/q} = x^{3}f^{2}     \]
Thus we have no dependence on $p$ and $q$ as far as the quasimonomials are
concerned.
The GLV matrices are:
\[ A = \left( \begin{array}{ccccc}
            -a  & -b  & 0  &    0     &  0      \\
            -pa & -pb & qa & q(b+2da) & 2bdq
              \end{array}   \right)  \; ,  \]
\[ B = \left( \begin{array}{cc}
         -p/q   & 1/q \\
         1-p/q  & 1/q \\
         1-2p/q & 2/q \\
         2-2p/q & 2/q \\
         3-2p/q & 2/q
       \end{array} \right) \; , \;\:
\lambda = \left( \begin{array}{c}  0 \\ 0  \end{array}  \right)   \]
And the LV matrices are given by:
\[  \lambda ' = B \cdot \lambda = \left( \begin{array}{c}
                                   0 \\ 0 \\ 0 \\ 0 \\ 0
                                  \end{array}  \right) \: , \;\:
    A ' = B \cdot A = \left( \begin{array}{ccccc}
              0  &  0  & a  &   b+2da  & 2bd  \\
             -a  & -b  & a  &   b+2da  & 2bd  \\
             -a  & -b  & 2a & 2(b+2da) & 4bd  \\
             -2a & -2b & 2a & 2(b+2da) & 4bd  \\
             -3a & -3b & 2a & 2(b+2da) & 4bd
                       \end{array} \right)              \]
Thus the LV matrices do not depend on $p$ and $q$, just like the
quasimonomials. The LV system is $(p,q)$--independent.

As a second example we shall mention the Morse oscillator, of
relevance in the field of molecular structure~\cite{b&j}:
\begin{equation}
      \ddot{x} = -2d\alpha e^{-\alpha x} (1 - e^{-\alpha x})
\end{equation}
Setting $y = \dot{x}$ we arrive at a first order differential system:
\begin{eqnarray*}
\dot{x} & = & y  \\
\dot{y} & = & -2d\alpha e^{-\alpha x} (1 - e^{-\alpha x})
\end{eqnarray*}
This system has the form (\ref{eq:ini}). The obvious choices at this stage
are:
\[  f(x) = e^{-\alpha x} \;\: , \frac{\mbox{d}f}{\mbox{d}x} = -\alpha f \]
Although $f(x) > 0$ for all $x$, this is not the case for $x$ and $y$. Thus
we must perform a phase-space translation of magnitude $c$, with $c$ large
enough to
ensure the positiveness of both $x$ and $y$. When this is done, followed by
the introduction of a new variable $z = x^{p} y^{p'} f^{q}$, the result is a
GLV system of strictly positive variables:
\begin{eqnarray*}
\dot{x} & = & x \left[ x^{-1}y - cx^{-1}\right]  \\
\dot{y} & = & y \left[ ax^{-p/q}y^{-1-p'/q}z^{1/q} - abx^{-2p/q}y^{-1-2p'/q}z^{2/q} \right] \\
\dot{z} & = & z \left[ \alpha cq + px^{-1}y - cpx^{-1} + ap'x^{-p/q}y^{-1-p'/q}z^{1/q} - \right.  \\
        &   & \left. abp'x^{-2p/q}y^{-1-2p'/q}z^{2/q} - \alpha qy \right]
\end{eqnarray*}
where $a = -2db\alpha $ and $b = e^{\alpha c}$. There are 5 quasimonomials,
which will be the variables of the resulting $5 \times 5$ LV system:
\[ x^{-1}y \; , \;\: x^{-1} \; , \;\: x^{-p/q}y^{-1-p'/q}z^{1/q} = y^{-1}f \; , \;\:
   x^{-2p/q}y^{-1-2p'/q}z^{2/q} = y^{-1}f^{2} \; , \; \: y             \]
The resultant GLV matrices are:
\[ A = \left( \begin{array}{ccccc}
             1  & -c  & 0   &    0     &  0            \\
             0  &  0  & a   &   -ab    &  0            \\
             p  & -cp & ap' &  -abp'   &  -\alpha q
              \end{array}   \right)  \; ,  \]
\[ B = \left( \begin{array}{ccc}
         -1    & 1        & 0      \\
         -1    & 0        & 0      \\
         -p/q  & -1-p'/q  & 1/q    \\
         -2p/q & -1-2p'/q & 2/q    \\
          0    & 1        & 0
       \end{array} \right) \; , \;\:
\lambda = \left( \begin{array}{c}  0 \\ 0 \\ \alpha cq \end{array}  \right)   \]
And the LV matrices will be now:
\[  \lambda ' = B \cdot \lambda = \left( \begin{array}{c}
                                   0 \\ 0 \\ \alpha c \\ 2\alpha c \\ 0
                                  \end{array}  \right) \: , \;\:
    A ' = B \cdot A = \left( \begin{array}{ccccc}
             -1  &  c  &  a  & -ab &   0       \\
             -1  &  c  &  0  &   0 &   0       \\
              0  &  0  & -a  &  ab & -\alpha   \\
              0  &  0  & -a  &  ab & -2\alpha  \\
              0  &  0  &  a  & -ab &   0
                       \end{array} \right)              \]
Again, both the LV variables and matrices are independent of $p$, $p'$
and $q$.

\mbox{}

\mbox{}

\begin{flushleft}
We would like to thank Dr. L. Brenig for fruitful discussions and suggestions.
\end{flushleft}

\pagebreak

\end{document}